\documentclass[preprint,showpacs,preprintnumbers,amsmath,amssymb]{revtex4}

\usepackage{graphicx}% Include figure files
\usepackage{dcolumn}% Align table columns on decimal point
\usepackage{bm}% bold math

\begin{document}

\preprint{APS/123-QED}

\title{First-Principles Study on Peierls Instability in Infinite Single-Row Al Wires}

\author{Tomoya Ono}
%\email{ono@upst.eng.osaka-u.ac.jp}
\affiliation{Research Center for Ultra-Precision Science and Technology, Osaka University, Suita, Osaka 565-0871, Japan}

\author{Kikuji Hirose}
\affiliation{
Department of Precision Science and Technology, Osaka University, Suita, Osaka 565-0871, Japan}

\date{\today}% It is always \today, today,

\begin{abstract}
We present the relation between the atomic and spin-electronic structures of infinite single-row atomic wires made of Al atoms during their elongation using first-principles molecular-dynamics simulations. Our study reveals that the Peierls transition indeed occurs in the wire with magnetic ordering: it ruptures to form a trimerized structure with antiferromagnetic ordering and changes from a conductor to an insulator just before forming a linear wire of equally-spaced atoms. The formation of the trimerized wire is discussed in terms of the behavior of the $\sigma$-symmetry bands of the Al wire.
\end{abstract}

\maketitle
In the past decade, there has been a great deal of interest in the preparation of atomic wires and the investigation of their mechanical and electrical behaviors \cite{datta-ruitenbeek}. The properties of such minute systems with low dimensionality can be different from those of the bulk. In particular, an infinite single-row atomic wire (ISAW) has attracted great attention from a fundamental point of view, because phenomena peculiar to the reduced dimensionality are expected to emerge best for the ISAW having an ultimately simplified one-dimensional structure \cite{portal-torres-maria-ono,watanabe,ayuela,sen,ono}. Half of a century ago, Fr\"ohlich \cite{frohlich} and Peierls \cite{peierls} discovered instability in one-dimensional electron systems that has come to be called the Peierls instability: the regular chain structure in a one-dimensional wire with a partly filled band will never be stable, since one can always find a more suitable structure, such as a dimerized structure, for which a break will occur at or near the edge of the Fermi distribution. Thus the band will be split into a number of smaller zones, some of which are filled and the rest empty, and as a consequence, the wire changes from a conductor to an insulator, which is the so-called Peierls transition. So far, several quasi-one-dimensional metals such as organic ones were observed to exhibit the Peierls transition \cite{claessen}.

First-principles molecular-dynamics (FPMD) simulations have been employed to elucidate the atomic structures and spin-electronic properties of atomic wires \cite{portal-torres-maria-ono,ayuela,sen,ono}. The studies for an ISAW made of {\it gold} atoms \cite{portal-torres-maria-ono,ono} demonstrated that the Peierls transition certainly occurs: the gold wire breaks to form a dimerized structure accompanied with a metal-insulator transition when the average interatomic distance increases beyond $\sim$ 5.7 a.u., where a zigzag geometry transforms into a linear one in the process of its elongation. What has to be noted is that a gold atom is considered simple due to its $s$-type outermost-valence orbital and its nonmagnetic property. The situation may be more complex in the case of monatomic wires made of elements having valence orbitals other than the $s$-type one and/or made of magnetic elements. Recently, for an ISAW consisting of {\it aluminum} atoms (Al-ISAW) with 3$s$- and 3$p$-type valence orbitals, Ayuela {\it et al.} \cite{ayuela} reported that an ISAW with a regular chain structure of equally-spaced atoms on a straight line is in a metallic state with certain magnetic ordering and it is energetically more favorable than a dimerized wire at an interatomic distance of 5.3 a.u., which is close to the nearest-neighbor atomic distance of 5.4 a.u. in the aluminum bulk. Thus, the existence of a metallic linear ISAW of equally-spaced atoms gives rise to a puzzle in Peierls's view that such an atomic wire will never be stable, although the result is appealing in that a stable ISAW is in a magnetic state. Another theoretical study by Sen {\it et al.} \cite{sen} suggested the possibility of the Al-ISAW to rupture into a tetramized wire under the assumption that the effect of the distortion on cohesive energy is negligibly small, but they overlooked the occurrence of magnetic ordering in the wire. Therefore it is imperative to clarify the properties of the Al-ISAW for understanding of the Peierls instability in a one-dimensional atomic wire during its elongation.

In this Letter, we implement detailed FPMD simulations to elucidate the close relation between the atomic and spin-electronic structures of the Al-ISAW. Our finding is that the Peierls transition indeed occurs in the Al-ISAW with magnetic ordering: the ISAW during its elongation breaks just before it transforms from a zigzag structure to a linear one. When the wire fractures, the spin-density wave (SDW) arising from the Coulomb repulsion interaction leads the wire to spontaneously stabilize, and opens up an energy band gap at the Fermi level ($E_F$). Consequently, a trimerized wire with antiferromagnetic (AFM) ordering emerges, i.e., the wire transforms from a conductor into an insulator at the formation of the trimerized wire with AFM ordering. Moreover, we confirm that this distorted structure is dominated by the behavior of the $\sigma$-symmetry bands crossing the $E_F$.
\begin{figure}[bt]
\includegraphics{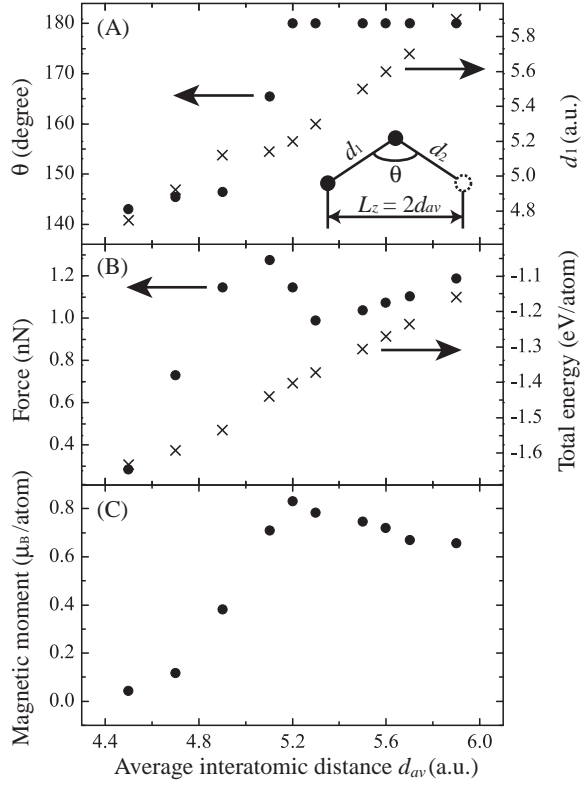}
\caption{(A) Bond angle $\theta$ (closed circles) and interatomic distances $d_1$ (crosses), (B) restoring force (closed circles) and total energy per atom (crosses), (C) magnetic moment for the model of the Al-ISAW as shown in the inset of (A). The supercell of $N_{atom}$=2 is employed. The zero of energy is chosen to be the total energy of an isolated aluminum atom. In the inset of (A), the closed circles denote the atoms in a supercell and the open circle shows the replicated atom in the adjacent supercell.}
\label{fig:1}
\end{figure}

Our FPMD simulation is based on the real-space finite-difference method \cite{rsfd}, which enables us to determine the self-consistent electronic ground state and the optimized atomic geometry with a high degree of accuracy by making use of the timesaving double-grid technique \cite{tsdg} and the direct minimization of the energy functional \cite{dmef}. The norm-conserving pseudopotentials \cite{ncps} of Troullier and Martins \cite{tmpp} are employed to describe the electron-ion interaction. Exchange-correlation effects are treated with the generalized-gradient approximation \cite{gga} in the density-functional theory. The nine-point finite-difference formula (i.e., the formula with $N$=4) \cite{rsfd} is adopted for the derivative arising from the kinetic-energy operator. We take a cutoff energy of 27 Ry, which corresponds to a grid spacing of 0.6 a.u., and a higher cutoff energy of 247 Ry in the vicinity of the nuclei with the augmentation of double-grid points \cite{tsdg}. We employ a technique that involves the use of a periodic supercell whose size is chosen as $L_x$=$L_y$=20 a.u. and $L_z$=$N_{atom} \times d_{av}$. Here, the $z$ axis is taken along the wire axis, the $x, y$ axes are perpendicular to the wire, $L_x$, $L_y$ and $L_z$ are the lengths of the supercell in $x$, $y$ and $z$ directions, respectively, $N_{atom}$ is the number of aluminum atoms within the supercell, and $d_{av}$ is the average value for {\it projections} of the interatomic distances between adjacent atoms onto the $z$ component. In order to eliminate completely unfavorable effects of atoms in neighbor cells which are artificially repeated in the case of periodic boundary condition, we impose the {\it nonperiodic} boundary condition of vanishing wave function out of the supercell in the $x$ and $y$ directions and adopt the periodic boundary condition in the $z$ direction.
\begin{figure}[bt]
\includegraphics{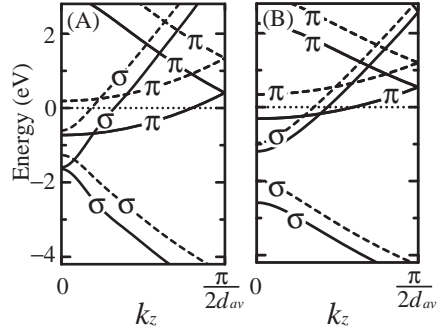}
\caption{Energy band structures of the Al-ISAW at (A) $d_{av}$=5.2 a.u. and (B) $d_{av}$=5.9 a.u. plotted along the direction $k_z$ in reciprocal space. The supercell of $N_{atom}$=2 is employed. The zero of energy is chosen to the Fermi level $E_F$. Solid and dashed curves represent up-spin and down-spin electron bands, respectively. The $\pi$-symmetry bands are doubly degenerate.}
\label{fig:2}
\end{figure}

First we examine the stability of the Al-ISAW in the case of $N_{atom}$=2. The calculation model, the parameters of the fully optimized geometry of the ISAW, the restoring force, the total energy per atom, and the magnetic moments are depicted as a function of $d_{av}$ in Fig.~\ref{fig:1}, and the energy band structures of the ISAW are plotted in Fig.~\ref{fig:2}. In these calculations, 24 $k_z$ points are employed for the integration over Brillouin zone. As an initial configuration for FPMD calculation, the parameters of the atomic geometry $d_1$ and $\theta$ at $d_{av}$=4.45 a.u. are randomly set in the ranges of 4.45$\pm$0.05 a.u. and 140$\pm$5 degrees, respectively \cite{comment1}. Then, we increase $d_{av}$ and relieve the force on atoms repetitiously. In each case, the structural optimization is performed until the remaining forces acting on the atom are smaller than 82.4 pN. From Figs.~\ref{fig:1} (A) and ~\ref{fig:2}, one sees that a zigzag wire changes to a linear one upon elongating $d_{av}$ more than $\sim$ 5.2 a.u., and even when further stretching up to $d_{av}$=5.9 a.u., the wire does not form a dimer structure, but keeps a regular chain structure of equally-spaced atoms on a straight line \cite{comment2}, and it remains a conductor. These results throw us into a {\it puzzling} in the view of the Peierls instability. The wire in ferromagnetic (FM) state has been also observed to be energetically more stable than those in other magnetic states, i.e., AFM and paramagnetic (PM) orderings, over the range of $d_{av}$ calculated here in the case of $N_{atom}$=2.
\begin{figure}[bt]
\includegraphics{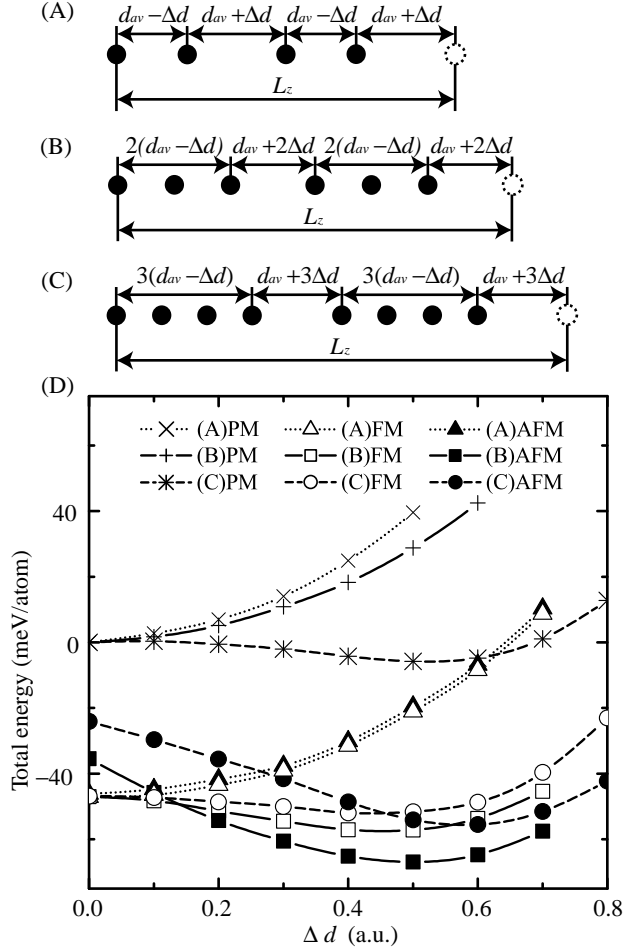}
\caption{Models for (A) dimerized wire of $N_{atom}=4$, (B) trimerized wire of $N_{atom}=6$, and (C) tetramized wire of $N_{atom}=8$. (D) Total energy per atom as a function of the bond length of the dimer in the wire with PM, FM, and AFM orderings. The zero of energy is taken to be the total energy of the wire with PM ordering at $\Delta d$=0 a.u. The average interatomic distance $d_{av}$ is set to be 5.6 a.u. In (A), (B), and (C), the closed circles denote the atoms in a supercell and the open circle shows the replicated atom in the adjacent supercell.}
\label{fig:3}
\end{figure}

To gain an insight into which structure is the most stable after the Al-ISAW breaks, we evaluate the total energies of the wires with several atomic and spin-electronic structures. Whereas the atoms in the wire were alowed only to dimerize in the above study, additional geometrical degrees of freedom, such as trimerization and tetramerization, are now taken into account for definite clarification of the atomic and spin-electronic structures of the Al-ISAW during its elongation. The calculation models and results are depicted in Fig.~\ref{fig:3}. We take the average interatomic distance $d_{av}$ of 5.6 a.u., which is longer than the nearest-neighbor atomic distance of 5.4 a.u. in the fcc aluminum crystal and than the critical distance of 5.2 a.u. where an ISAW transforms from a zigzag structure into a linear one. The Brillouin-zone integrations are performed using 12, 8, and 6 $k_z$ points for the models (A), (B), and (C) with supercells of $N_{atom}$=4, 6, and 8, respectively. In Fig.~\ref{fig:3} (B), the middle atoms within the trimers are located at the centers of the respective trimers in an intertrimer distance of 3$d_{av}$. In the case of the tetramized ISAW in Fig.~\ref{fig:3} (C), the two middle atoms inside the tetramers are relaxed through the FPMD simulation, while both the edge atoms of the tetramers are frozen so as to maintain a tetramer length of 3($d_{av}-\Delta d$) and an intertetramer distance of 4$d_{av}$. The electronic ground states with PM, FM and AFM orderings are observed, but that with the ferrimagnetic ordering is not found. In contrast to the conclusion drawn by Sen {\it et al.} \cite{sen}, we can find a sensible difference in energy between regular and distorted wires. The total energy per atom of the trimerized wire with AFM ordering is the lowest, as shown in Fig.~\ref{fig:3} (D). Therefore we conclude that the trimerized Al-ISAW in AFM state is the most likely structure for the distorted Al-ISAW.
\begin{figure}[bt]
\includegraphics{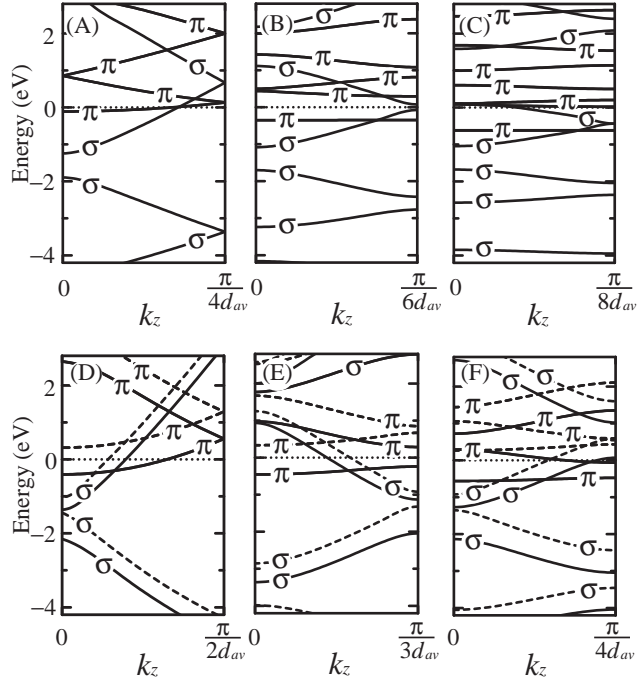}
\caption{Energy band structures of the Al-ISAW: (A) dimerized wire at $\Delta d$=0.0 a.u. with AFM ordering, (B) trimerized wire at $\Delta d$=0.5 a.u. with AFM ordering, (C) tetramized wire at $\Delta d$=0.6 a.u. with AFM ordering, (D) dimerized wire at $\Delta d$=0.0 a.u. with FM ordering, (E) trimerized wire at $\Delta d$=0.5 a.u. with FM ordering, and (F) tetramized wire at $\Delta d$=0.4 a.u. with FM ordering plotted along the direction $k_z$ in reciprocal space. Same as Fig.~\ref{fig:3} for the definition of $\Delta d$. In (D), (E), and (F), solid and dashed curves represent up-spin and down-spin electron bands, respectively. The zero of energy is taken to the Fermi level $E_F$. The average interatomic distance $d_{av}$ is set to be 5.6 a.u.}
\label{fig:4}
\end{figure}

We next consider why the Al-ISAW prefers a trimerized structure with AFM ordering after it breaks. This reason is understood by noting the behavior of the $\sigma$-symmetry band. In Fig.~\ref{fig:2}, one see that the stretched wire has doubly-degenerate $\pi$-symmetry bands and nondegenerate $\sigma$-symmetry ones near the Fermi level $E_F$, and the $\pi$-symmetry and $\sigma$-symmetry bands cross the $E_F$ near $k_z\approx\frac{\pi}{3d_{av}}$ and $k_z\approx\frac{\pi}{6d_{av}}$, respectively. When the ISAW ruptures, the SDW which has 6$d_{av}$ periodicity appears so as to open up an energy gap of $\sigma$-symmetry bands at the $E_F$ and to lower the total energy. Figure~\ref{fig:4} (B) shows the energy band structures of the trimerized ISAW with AFM ordering at $d_{av}$=5.6 a.u., for which one can find the energy band gap at the $E_F$. On the other hand, we can confirm that the SDWs which have the other periodicity, such as 2$d_{av}$ and 3$d_{av}$, lead the $\sigma$-symmetry bands to cross the $E_F$ and increase the total energy of the ISAW relative to the trimerized ISAW with AFM ordering having SDW of 6$d_{av}$ periodicity (see Fig.~\ref{fig:4}).
\begin{table}[bt]
\caption{Total energy per atom (eV/atom) of (A) the trimerized Al-ISAW with AFM ordering and (B) the Al-ISAW calculated using the model in Fig.~\ref{fig:1} (A). The zero of energy is chosen to be the total energy of an isolated aluminum atom.}
\begin{tabular}{c||cc}
\hline 
$d_{av}$ (a.u.)& \hspace{1cm}(A)\hspace{1cm} & \hspace{1cm}(B)\hspace{1cm} \\ \hline \hline
5.0 & $-$1.434 & $-$1.494 \\
5.1 & $-$1.410 & $-$1.422 \\
5.2 & $-$1.385 & $-$1.382 \\
5.3 & $-$1.359 & $-$1.350 \\
5.4 & $-$1.334 & $-$1.321 \\
\hline
\end{tabular}
\label{tbl:4}
\end{table}

Finally, we explore the critical interatomic distance where the Al-ISAW transforms from the zigzag structure to the trimerized one during its elongation. Table~\ref{tbl:4} shows the total energy per atom of the trimerized Al-ISAW with AFM ordering at several average interatomic distances $d_{av}$. The nearest-neighbor interatomic distance between the atoms in the trimers and the bond angle formed by two adjacent trimers are optimized until the maximum forces on the atom are smaller than 82.4 pN. The total energy of the wire for $N_{atom}$=2 given in Fig.~\ref{fig:1} is also depicted in the column (B) of Table~\ref{tbl:4} for comparison. From Table~\ref{tbl:4}, one can see that before $d_{av}$ increases up to 5.2 a.u., where the ISAW in Fig.~\ref{fig:1} has not changed to the linear wire yet, the ISAW forms a trimerized structure. We further comfirmed that the ISAW with AFM ordering is not a conductor but an insulator at $d_{av}$=5.2 a.u. Thus, the Al-ISAW exhibits the Peierls transition just before forming a linear wire in a metallic state.

In summary, we have studied the relation between the atomic and spin-electronic structures of the Al-ISAW using FPMD simulations. We have observed that the Peierls transition occurs in the Al-ISAW with magnetic ordering: the Al-ISAW has a zigzag structure in a FM state below $d_{av}$=5.2 a.u. Just before the average interatomic distance increases to 5.2 a.u., the wire ruptures to form a trimarized structure with AFM ordering. The Al-ISAW changes from a conductor to an insulator upon breaking. The preference for a trimerized structure with AFM ordering in the distorted wire is attributed to the nature of its $\sigma$-symmetry bands.

This research was partially supported by the Ministry of Education, Culture, Sports, Science and Technology, Grant-in-Aid for Young Scientists (B), 14750022, 2002. The numerical calculation was carried out by the computer facilities at the Institute for Solid State Physics at the University of Tokyo, and Okazaki National Institute.


\begin{references}
\bibitem{comment1}
In the recent comparative study attempting to understand the stability of a planer chains in zigzag structure \cite{sen}, the energy shows a shallow minimum at $d_{av}$=4.48 a.u., with a bond angle of 139$^\circ$. This geometry is employed as an initial configuration for our FPMD simulations.
\bibitem{comment2}
Since a linear ISAW of equally-spaced atoms emerges in the range of $d_{av} \geq$ 5.2 a.u., the band structure in the case of $N_{atom}$=1 is the same as that in Fig.~\ref{fig:2}.
\end{references}
\end{document}